# Collective Innovation in Open Source Hardware

HARRIS KYRIAKOU AND JEFFREY V. NICKERSON, Stevens Institute of Technology

## 1. INTRODUCTION

A growing community that shares digital 3D designs has created an opportunity to study, encourage and stimulate innovation. This remix community allows people not only to prototype at a minimal cost but also to work on projects they are genuinely interested in (Acosta 2009; Dougherty 2012). Participants free of the limitations typically imposed by formal organizations develop products driven by their own interest.

The traces of this activity can be studied in the hopes of understanding what generates new inventions. At of the heart of this activity is remixing, the process through which community members modify or combine each other's ideas. Such a remixing process, under different names, has been seen as an integral part of innovation for many decades (van den Bergh 2008; Kogut and Zander 1996; Schumpeter 1934). However, research on the actual mechanics of such processes has been limited (Benner and Tushman 2003).

Potential focal points for such research include understanding better the role of modularity in remixing (Fleming and Sorenson 2004), as well as understanding design diversity (van den Bergh 2008), related to the expertise and leadership of the designers. The ways designers combine each other's ideas, the aspects of each design they inherit, the role critique plays – all are topics of study that could lead to creativity support systems in support of product innovation.

The focal point of our research is Thingiverse, an open and dynamic community of designers where participants can freely download, modify or combine designs (Figure 1). Thingiverse creations are intended for 3D printing and vary in levels of complexity and applicability, ranging from toys and mugs to robots and quadcopters. Thingiverse makes it possible to observe novice users acquiring tacit knowledge while interacting with experts (Miller et al. 2006). It is also possible to see how and how often the ideas are implemented (Baer 2012).

## 2. A NOVELTY METRIC

Product innovation is defined in terms of the change in a product's design (Katila and Ahuja 2002). Preceding studies have measured innovation, originality and creativity through patent networks (Ahuja 2000; Murray 2002; Schilling and Phelps 2007). Ahuja (2002) recognized the limitations of using archived patent data – in particular, the citations lag by years – and suggested the need for complementary approaches. Other methods used include self-reports (Axtell et al. 2000; Baer 2012; Shalley et al. 2009) and ranking methods where consumers classify products along many dimensions (Von Hippel 1986). There is also a social aspect to innovation, leading to analysis of innovator networks (Tuomi 2002).

In the case of Thingiverse, participants link their designs to parent designs, creating a design inheritance network. Essentially, they remix other CAD files, drawing upon the work of others in order to produce new work (Lessig 2008). Their links are direct acknowledgement of the information they have used (for more on remixing, see Cheliotis and Yew, 2009).

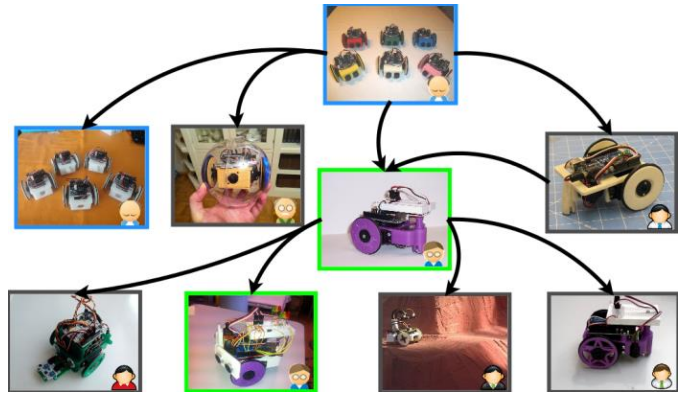

Fig. 1. Inheritance Network of robot designs shared, improved and remixed in Thingiverse.

In our own work, we have analyzed this design inheritance network of over 16,000 designs. In addition, we quantified design similarity by incorporating a shape comparison method from the computer graphics literature (Kazhdan et al. 2003) modified to function with 3D printing files and the multiple components nature of some designs (Kyriakou and Nickerson 2013).

Such a method allows us to measure the distance of each newly submitted design to its closest neighbor among preexisting designs in a high dimensional space. Any design space can be visualized as a landscape by using multidimensional scaling to retain two distance dimensions in the plane and adding a vertical dimension to indicate some measure of value: for example, popularity (Figure 2). Such a representation of users' collective effort to innovate can provide additional insights, consistent with Fleming and Sorenson's (2001) view of invention as a process of recombinant search.

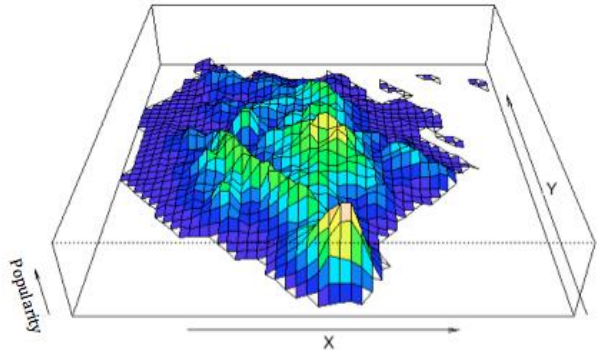

Fig. 2 Multidimensional scaling of designs represented in a multidimensional space. Z-axis represents popularity of a design.

## 3.   OPEN & COLLECTIVE INNOVATION

The innovation literature has identified the not-invented-here syndrome (Katz and Allen 1982)**,** the tendency of people to reinvent rather than invent on top of others' work. Is this syndrome lessened or eliminated by using remix networks (Gupta et al. 2006; Kyriakou et al. 2012)? From the patent literature, and from other theoretical analysis we expect that remixing communities will create a norm of reuse, and that remixed designs will generate interest from other users (Fleming et al. 2007; Wang et al. 2010; Yu and Nickerson 2011).

How though, are we to represent such communities? One possible conceptualization focuses on the product network as the core. The products are linked to each other by inheritance links. The products also belong to a semantic network, the ontology of known product descriptions. Thus, each product has a link to one or many nodes in such an ontology. For example, patents are classified by patent offices in this way. The products are also linked to individual inventors. These inventors may be linked to each other, either implicitly through the product network, based on what designs each designer has remixed, or explicitly through the social networks enabled by the communities. Figure 3 shows this conceptualization.

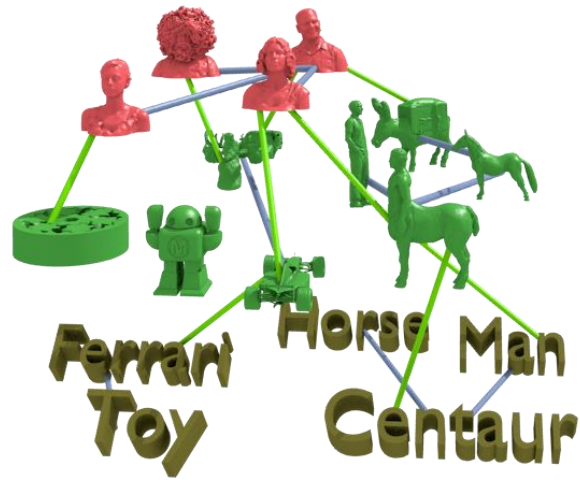

Fig. 3. The social network, on top, connected to the design inheritance network connected to the semantic network below.

## 4. GOING FORWARD

Remix networks represent a rich source of data for the understanding of collective innovation. Traces of many networks are shown, and, in particular, the modification and combination of ideas is shown explicitly. It becomes possible to understand the way designs evolve. This understanding, in turn, may lead to creativity support aids. That is, if we understand how inventors choose ideas to remix and combine, and we know which sets of decisions have done better in the past, it might be possible to suggest to inventors particular parts of the design space to explore. Such suggestions don't have to work all the time; they just need to work sometimes, nudging someone into a new part of the landscape, one that might twist up through the mist.

**ACKNOWLEDGEMENTS**: This material is based upon work supported by the National Science Foundation, grant number IIS-0968561.

4  H. Kyriakou and J.V. Nickerson